# Towards establishing formal verification and inductive code synthesis in the PLC domain

Matthias Weiß
Institute of Industrial Automation and
Software Engineering
University of Stuttgart
Stuttgart, Germany
ORCID: 0000-0002-5594-4661

Philipp Marks[1]
Institute of Industrial Automation and
Software Engineering
University of Stuttgart
Stuttgart, Germany
ORCID: 0000-0003-0531-4984

Benjamin Maschler[1]
Institute of Industrial Automation and
Software Engineering
University of Stuttgart
Stuttgart, Germany
ORCID: 0000-0001-6539-3173

Dustin White[1]
Institute of Industrial Automation and
Software Engineering
University of Stuttgart
Stuttgart, Germany
ORCID: 0000-0002-7321-0637

Pascal Kesseli
DiffBlue Ltd.
Oxford, United Kingdom
ORCID: 0000-0003-0300-5598

Michael Weyrich
Institute of Industrial Automation and
Software Engineering
University of Stuttgart
Stuttgart, Germany
ORCID: 0000-0003-3176-9288

*Abstract*— Nowadays, formal methods are used in various areas for the verification of programs or for code generation from models in order to increase the quality of software and to reduce costs. However, there are still fields in which formal methods haven't been widely adopted, despite the large set of possible benefits offered. This is the case for the area of programmable logic controllers (PLC). This article aims to evaluate the potential of formal methods in the context of PLC development. For this purpose, the general concepts of formal methods are first introduced and then transferred to the PLC area, resulting in an engineering-oriented description of the technology that is based on common concepts from PLC development. Based on this description, PLC professionals with varying degrees of experience were interviewed for their perspective on the topic and to identify possible use cases within the PLC domain. The survey results indicate the technology's high potential in the PLC area, either as a tool to directly support the developer or as a key element within a model-based systems engineering toolchain. The evaluation of the survey results is performed with the aid of a demo application that communicates with the Totally Integrated Automation Portal from Siemens and generates programs via Fastsynth, a model-based open source code generator. Benchmarks based on an industry-related PLC project show satisfactory synthesis times and a successful integration into the workflow of a PLC developer.

## I. INTRODUCTION

Programmable Logic Controllers (PLC) are the most commonly used type of controller when it comes to the control of manufacturing processes or industrial automation systems [1]. Having their origin in the 1960s, various programming languages with different abstraction level (graphical to object-oriented) [2] have been developed over the decades and can partly be used interchangeably. The most common PLC programming languages are standardized in IEC 61131-3 [3]. Legacy PLC code is often hard to understand which leads to large efforts and a high risk of errors when making changes to an existing program [4].

Although some evolution in the programming of PLCs can be seen, many concepts of modern software engineering that aim to increase the quality of software or to reduce its lifecycle cost have not been investigated in the context of PLCs yet. One of these concepts is the application of formal methods [5], e.g. formal verification, which is state of the art for verifying the correctness of safety-critical systems in other areas [6, 7].

One reason why formal methods have not yet been widely applied in the PLC domain may lie in the major gap between the theoretical concepts of formal methods originating from computer science and the PLC practitioners which usually have a technical background. The identified research gap that is addressed in this paper can be described by the following two aspects:

1. An engineering-oriented description of formal methods is missing which gives PLC practitioners an idea on how the technology works and how it can be used beneficially.
2. An integration of suitable tools that apply formal methods in the existing toolchain and workflow of PLC engineers is missing.

These aspects are considered as crucial factors by the authors that need to be addressed before formal methods can be beneficially applied in the development of PLC software. Thus, this paper tries to bridge the identified gap between computer science and engineering by considering both aspects.

The rest of this paper is organized as follows: Section II gives a short introduction on model-based development and the current use of formal methods in the PLC domain. Furthermore, this section introduces formal methods and gives an overview on existing tools. Section III describes our approach to identify and bridge the gap between formal methods research and PLC practitioners by applying the concept of "self-configuring blocks". Furthermore, the section describes the results of an industry survey on the understandability and potentials of this technology which revealed further use cases for applying formal methods to PLC programs in the future. Subsequently, Section IV describes the implementation details of integrating formal methods in the workflow of PLC developers. To achieve this, the Totally Integrated Automation (TIA) Portal by Siemens was extended to automatically generate PLC code by using an interface. Using this implementation, Section V presents the results of the evaluation in a small industrial scenario. Finally, in Section VI, the paper closes with conclusions and an outlook on future work.

---
[1] These authors contributed equally to this publication.

## II. STATE OF THE ART

### A. Model-based Development and Formal Methods in the PLC Domain

Model-based engineering describes an engineering process with models as main development artifacts instead of documents [8]. Model-based software engineering translates this concept into its own domain, and model-based systems engineering integrates it into a consistent and interdisciplinary black box and top-down approach [9]. Model-driven software engineering further extends on that: Model-driven means that models are the main artifacts and that these are linked to and can be transformed into each other over the whole development respectively engineering process [10]. For software engineering, the main implementation of this is code generation from models [10]. The concepts of model-based and model-driven engineering are widely, but not consistently implemented in the industry [11].

One reason for a slow adoption by enterprises is the initial implementation effort as well as the acquisition of a new respectively the change of an existing toolchain. Traditionally, for developing a program, at least an editor, a compiler, a linker, the runtime environment or hardware and a debugger are needed. In the case of model-driven software engineering, a model editor substitutes or extends the former editor [12]. The compiler is integrated and extended by a code generator that additionally translates a model into code within a model-driven software development toolchain [12]. Furthermore, depending on the abstraction level, either a hardware support package or an emulation of the runtime environment is needed to assist with the generation of specific and correct code that can run instantly on the target environment.

Formal methods are machine-readable techniques, mostly mathematical, to automate, analyze or verify development steps, models, or artifacts [13]. Formal methods can be applied in two ways: Firstly, in forward direction they aid in automatically generating correct artifacts for a derived model which is the main technique used for model-driven engineering. Secondly, they can be used backwardly to validate or verify new functions or properties against requirements or approved development artifacts. This is then called formal verification.

When models are used as main artifacts within the development process, formal methods can and should be applied to ensure consistency across the different models [14]. For example, any development process for safety critical systems in the automotive industry is required to integrate formal methods [15, 16].

In the PLC-domain, modelling the needed code via instruction list, structured text, ladder diagram, functional block diagrams or sequential function chart is state-of-the-art [17]. One prominent example of a tool supporting the complete model-based engineering workflow for PLC programs is the development platform TIA Portal by Siemens. To maintain a consistent model-driven development process from modelling to the automatically generated, executable program, formal verification would be helpful to automatically ensure the accuracy of the developed models and, thereby, of the generated code. There are several publications which explore the usage of formal methods in the context of the PLC domain: First methods that enable formal verification of PLC code were already described in the early 1990s [18]. Since then, researchers proposed a wide variety of approaches to make formal methods more accessible for PLC engineers, including domain-specific languages that support the creation of formal specifications [19]. In more recent years, concepts for PLC code generation based on formal methods were introduced as well. One example is a PLC-specific language, which can be used to generate code via control flow graphs [20].

Nevertheless, formal methods and especially formal verification and code generation are not widely used in industry [17, 21]. Having mentioned the slow adaption of model-driven engineering, we identify that both tool support and the integration in existing industry processes are two main obstacles for the establishment of formal methods in the PLC domain. One way to address those issues is an integration of formal methods into a well-established engineering tool and, thereby, the current engineering workflow.

### B. Tools for Practicing Formal Methods

Countless tools for the formal verification of programs have been developed, most of them targeted at either regular applications or embedded and safety-critical systems [22]. One of these tools is the C Bounded Model Checker (CBMC) that is being developed by Diffblue Ltd. and a small open source community. Its code is publicly available via GitHub (see [23]).

If a program is given to CBMC, it is first being transformed into an intermediate language (the so-called "GOTO language"), unwinding loops in the process if possible [24]. The resulting mathematical equations can then be passed to a decision procedure like a Satisfiability (SAT) solver in order to find any violations to the given constraints (such as assertions). Languages other than C can be supported by creating additional parser interfaces for the conversion into the GOTO language. So far, there exist interfaces for Verilog, Java (not open source) and the Siemens Statement List (STL) language with limited functionality. Since most other PLC languages can be expressed by using STL statements, this enables de-facto-compatibility with ladder and function block diagrams as well [25].

Model checkers can be used as a component in the Counterexample Guided Inductive Synthesis (CEGIS) algorithm in order to generate code [26]. For this, a formal specification that describes the properties of the desired program is needed. The algorithm starts by creating a program candidate that is syntactically valid. This program is then verified against the given specification by a model checker. If a counterexample that violates the specification is found, it is handed back to the synthesis procedure. This time, the synthesis tool generates a program that is both syntactically valid and fulfills all counterexamples that were found up to this point. The new candidate is then again being verified, effectively creating a loop between the synthesis and verification phase. This way, the procedure converges to a suitable program candidate and eventually finds it, if the given specification is satisfiable. "Simple" programs, meaning programs that contain fewer statements to solve the given task, are found first and thus preferred in the process. More details about the algorithm can be found in [26].

To our knowledge, there are no practical tools that make use of the CEGIS algorithm besides Fastsynth, whose GitHub project can be found at [27]. Fastsynth makes use of CBMC and is designed to meet four general use cases:

- **Code generation:** Creation of a program that meets a specification given by the user.
- **Code verification:** Automated testing of a program for certain properties with the aim of error detection and consistency checking.
- **Code repair:** Fixing bugs of a given program based on a specification and the source code of the original program.
- **Code simplification:** Transformation of a given program into the simplest possible representation without changing its externally visible behavior. The source code of the program is used as an input.

Along with CBMC, Fastsynth has already been extended by a PLC language interface. However, little effort has been made yet to determine whether the listed use cases are suitable for the PLC domain and what else could be accomplished by using the technology. This paper aims to expand on the pre-existing use cases by both searching for new use cases and transferring the new and known use cases into the PLC domain. The resulting concept is then being evaluated in an industrial application.

### III. EXPERT INTERVIEWS

In order to look for new use cases for formal verification and code generation in the PLC domain, a survey among PLC experts was conducted. Because the technology in question is largely unknown to this peer group, it was necessary to design a new notation that is suitable for PLC engineers beforehand. This section first covers the creation process of this description and its evaluation by the experts (subsection III.A). Subsequently, the methods and results of the expert interviews, which include both the extraction of new use cases (subsection III.B) and their assessed potentials (subsection III.C) are described.

#### A. Technology Transfer

In order to be able to analyze the potentials of formal methods inside the PLC domain, the technology first needs to be translated into a notation suitable for PLC engineers. This description served as a basis for this survey and was modified later in the process in order to reflect the feedback given by interview participants. The following principles were applied during the initial transformation:

*a) Familiarity:* Since the goal of the new presentation is to explain the technological concept to PLC engineers, it is beneficial to utilize terms and concepts that are known to this group. This allows professionals to communicate by using their own terminology and makes the technology more accessible. The same applies to images showing the technology inside the engineer's regular working environment.

*b) Simplicity:* The participants of the survey come from different domains and their experience in the field varies (see section B). Because each survey participant should be able to comprehend the technology description, the used language needs to be as simple as possible. Additionally, detailed explanations of terms and concepts originating from the domain of formal methods should be avoided wherever possible. At the same time, this must not affect the assessment of the potential.

*c) Extendibility:* It should be possible to easily modify the description based on the results of the survey. Finding new use cases for the technology should not result in fundamental changes of the concept.

Under consideration of these aspects, a new description for the technology was prepared. Since this new description should target PLC professionals, terms and concepts are used which are known to this group. One way to achieve this is to make use of the structure of a PLC project: PLC projects consist of blocks that communicate with each other via their interfaces (similar to e.g. functions and objects in classic programming languages). There are several different block types that differ in their core features. For example, some blocks are able to save data between block calls while others are not.

Based on this knowledge, a concept was created that wraps the key functionalities of the verification and generation tools mentioned in subsection II.B into a new type of block. These were called "self-configuring blocks" and described by the following properties:

- Just as most other block types, self-configuring blocks have an interface in which the user is able to declare new variables that can be used during the code generation or verification process.
- As the name suggests, self-configuring blocks are not directly programmed by the user. Instead, they contain a so-called "constraint list" that is configurable by the user. This list includes conditions that should always be satisfied by the block.
- It is possible to switch between use cases by selecting a different mode from a drop-down menu. The first draft of the description includes all basic use cases mentioned in subsection II.B.

Self-configuring blocks still contain a section with code and networks, showing the generated code and the block's exact behavior in the classical way. This code can also be modified if the verification use case is selected, allowing for the verification of pre-existing program code.

After creating this new description, it was evaluated by a total of ten professionals in the first part of the survey. Subject to investigation were its comprehensibility and the simplicity of newly introduced terms. To achieve this, a questionnaire with a total of four single-choice questions was designed. Common guidelines in the field of survey construction, such as the avoidance of suggestive questions, were considered during its creation process. For each question, the participants could select one out of five possible answers, ranging from "completely disagree" to "completely agree". Each participant had the opportunity to express additional feedback about the description and the technology itself. Fig. 1 shows the questionnaire's findings.

Overall, the technology description was received positively without any major comprehension problems. In general,

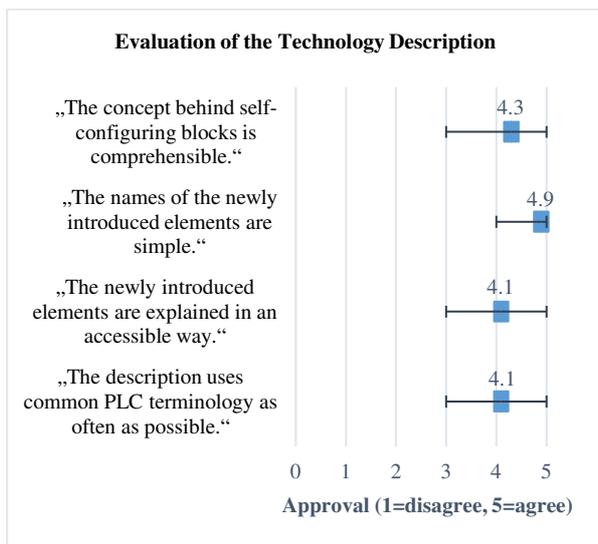

Fig. 1. Results of the survey about the technology description (n=10)

professionals with higher degrees of experience and those actively involved in the development process had the least difficulties in understanding the new technology. Two participants wished for a demo application and more details about the concept in order to make it more accessible. Still, the performed technology transfer can be considered a success and it is unlikely that the subsequent phases of the survey were impacted negatively by any comprehension problems.

*B. Extraction of Use Cases*

In order to evaluate the potential of formal methods in the PLC domain, the ten professionals were consulted again in the second phase of the survey. This phase aimed to, firstly, rate the potential of the technology based on the professionals' opinions and, secondly, identify previously unknown use cases for the technology in the PLC industry. For this purpose, the survey phase was split into two parts:

*a) Questionnaire about the potential of existing use cases:* Quantitative methods were used in order to determine how the participants would rate the potential of the technology and already existing use cases (described in section II) based on the provided description. For this, yet another questionnaire with five single-choice questions was designed. For each use case and the technology itself, the participants could rate their potential from "very low" to "very high".

*b) Oral questioning about potential use cases for the technology:* In order to extract use cases for formal methods in the PLC domain, the participants were questioned about different aspects of their profession. A semi-structured interview style was chosen as the strategy of this survey part for allowing participants to freely express their opinions and experiences while at the same time retaining a general interview structure. As a part of this approach, a pool of questions was designed to guide the interviewer. Each question was categorized based on its content, leading to the following theme blocks: 1) status of the professional, e.g. age, field and experience, 2) daily routine of a PLC engineer, 3) processes inside a PLC development project, 4) education of a PLC engineer, and 5) direct questions about new use cases or the enhancement of already existing use cases. Each block consisted of one to three key questions that were supported by several sub questions. This way, the interviewer could propose a key question to the participant and use the sub questions for further clarification of the topic when necessary. Additionally, block 5) was used between those questions as an attempt to connect the expressed experiences with the technology. All interviews were recorded and transcribed later. Those transcriptions served as the basis for the use case extraction, either by directly following the suggestions of the professionals or by further analyzing their statements, combining them with further research in the field.

Several new use cases could be identified by applying this method. These include either new presentation forms and sources for the constraint list or operations for the modification of existing code:

- **Code generation via truth tables:** Provide the possibility to display parts of the constraint list as a common truth table. The columns of the table include a selection of block interface variables, each row represents one constraint.

- **Code generation via "Cause and Effect"-Matrix**: Display parts of the constraint list as a matrix where the columns contain output and the rows contain input variables. The relationship between an output and one or more input variables can then be specified by filling up the cells of the corresponding column.

- **Generation via other models and process data**: Generation of PLC programs from already existing models and process data created during the development process. An example would be the generation and verification of code from a "Siemens Automation Designer" project.

- **Template-based generation:** Introduction of templates for known and often recurring problems, which can either be supplied with the product or created by the user. A template already contains the list of conditions (complete or incomplete). The code generation for the current project is done by selecting a suitable template and adjusting the condition list.

- **Code extension:** Add new behavior to existing blocks and networks by modifying the present code as little as possible.

- **Repair after analysis:** Suggest possible solutions when errors are found during code analysis and previously specified conditions are violated. Automatic adaptation of the code after the user has selected a solution.

- **Change language:** Conversion of the block into another programming language without changing the behavior that is visible from the outside.

*C. Assessment of New Use Cases' Potential*

After the new use cases were identified, a third survey phase, also known as member check, was initiated to validate their potential. For this, a new questionnaire which contained a description of the new use cases was designed. Again, participants were asked to assess the potential of the use cases from "very low" to "very high". A total of five professionals took part in this procedure. Fig. 2 shows the estimated potentials for both basic and new use cases.

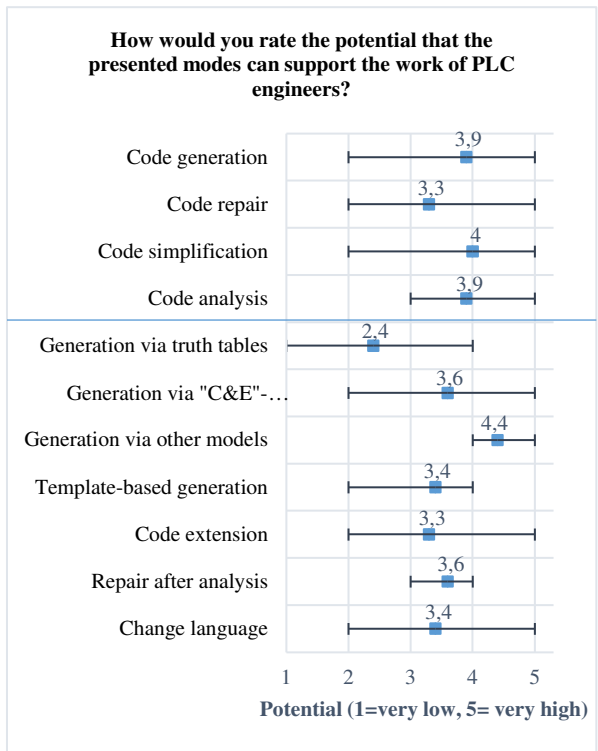

Fig. 2. Estimated potential of new (below blue line) and basic (above line) use cases for formal methods (New use cases: n=5; Basic use cases: n=10)

The potential of code generation and its characteristics was generally considered to be high (3.9). Two critics of this use case argued that the concept of the constraint list does not seem reusable enough. Especially in large projects, this would reduce the possible applications, since many different actors are involved. If, for example, a developer would want to use and modify the self-configuring blocks of another person, they would need to first understand the logical procedure of the other person in order to avoid undesired side effects. Another aspect that inhibits the potential is the length of the constraint list. Doubts have been expressed here by four participants as to whether modeling via the constraint list really proves to be faster than classical programming via the PLC languages.

This implies the reason why the potential of code generation from models and templates was rated generally higher than other generation use cases: By using templates, the constraint-based models become more controllable and their reusability increases. The use of other models simplifies the creation process of the constraint list or adopts it completely, leading to a significant decrease in development time.

The potential of code repair and extension is rather in the middle of the range (3.3) among the interviewees. This can be traced back to differences in the development process and the available tools. Some participants (n=2) could not imagine using the technology themselves, as they would fall back to alternative methods in conceivable application scenarios. Especially in larger projects, errors occurring during the test process are often more complex and cannot be eliminated by simply adding or removing a condition. Repair after code analysis partially automates the procedure for some errors and is therefore rated higher.

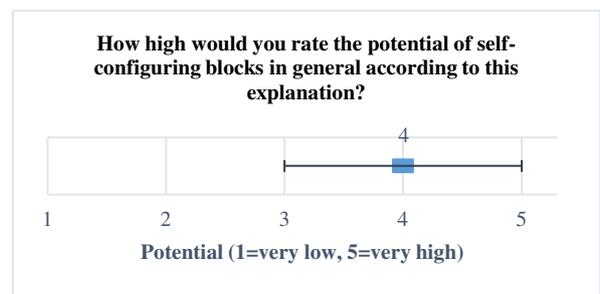

Fig. 3. Estimated general potential of self-configuring blocks (n=10)

The simplify use case was rated high (4.0) and particularly well received by engineers who often have to review and modify code from other developers in their daily work. Critics of the use case have doubts about whether the solution generated by the self-configuring blocks algorithm is actually easier to understand. A shorter solution would not automatically mean that the program code is easier for a human to read. This use case therefore faces the challenge of generating not only a short but also comprehensible program.

The potential of code analysis was rated high (3.9) by the participants with comparatively low variance. It was emphasized here that the technology is increasingly spreading in the PLC world, but still encounters acceptance problems in some domains. It should also be noted that other test techniques are used in practice in addition to code analysis, for example unit tests and simulations. The highest potential for this use case is therefore in the area of safety-critical systems where a safety case is necessary to minimize property damage or injuries.

Fig. 3 shows the estimated general potential of self-configuring blocks (n=10). The results imply that the interest in methods for automatic code generation is high across domains and that they offer the potential to support both companies and individual developers in various application scenarios. The trend towards formal methods is currently emerging, especially in the area of safety-critical systems, in order to shorten the test and acceptance process. However, due to the increasing use of models in e.g. plant automation there is an ever-greater need for compatible and reliable code generators. This is both an opportunity and a challenge for the technology. As the evaluation of the use cases already implies, each domain has its own requirements for such tools and a wide range of functions is essential for their establishment. In order to meet the needs of a domain, special solutions are required that are tailored towards the respective application scenario.

## IV. IMPLEMENTATION

The results of the survey were evaluated by using a demo application that implements some of the use cases found in section III. It was investigated whether formal methods could be integrated into the regular workflow of a PLC developer when creating or modifying the code of a PLC project. Since the PLC interface of the selected code synthesis tools (see section II.B) is designed for Siemens PLC languages, the workflow of engineers that are using PLCs by Siemens and the respective development environment, the Siemens Totally Integrated Automation (TIA) Portal, was considered.

The application aims to be as close as possible to the technology description used in the survey. Ideally the application should not interfere with the development in a way that would make it necessary for the user to adjust its everyday routines. It should provide additions to the current workflow while at the same time being simple to operate. Thus, the acceptance of the new technology should not be influenced negatively by e.g. a poorly designed user interface or the lack of automation for otherwise tedious tasks.

To achieve this, the coding interface of TIA Portal, the TIA Openness library, was used in order to communicate with the development environment of Siemens PLC engineers. Fig. 4 shows the described system architecture and how the application communicates with the environment.

With TIA Openness it becomes possible to load the data of existing PLC projects and to access and modify existing program code or to create completely new programs. This functionality can also be used in the code verification and generation by formal methods. After reading the user input, the application communicates with TIA Openness and the PLC interface of the tools presented in section II.B in order to achieve the given task. The results of the verification or code generation are then displayed by the application or imported in the PLC project.

Fig. 5 shows the main window of the program. By communicating with the TIA Portal, it is possible to load a PLC project and its software components which are then displayed on the left. If the user then selects a compatible block, a new XML file for the constraint list is created inside the PLC project or the already existing constraint list is loaded.

After the constraint list is loaded, it is displayed right from the project overview. The user can then modify the specification by adding, editing or removing constraints. For this first demonstration, the truth table code generation use case was implemented, enabling the possibility to generate programs consisting of Boolean IO mappings. Once the specification is complete, the user is able to start the process via the corresponding upper-right button. If a solution is found, the new code is automatically imported in the actual PLC project and can be viewed inside the TIA portal.

Overall, it was possible to integrate formal methods into the regular PLC developer workflow by automating most tasks and providing a user-friendly interface. Due to technical limitations of the TIA Openness interface however, it was not possible to integrate the concept of self-configuring blocks directly in the TIA Portal. A separate application is still needed for operation.

## V. EVALUATION

In order to evaluate whether the application is suitable for industrial scenarios, its performance and the quality of the retrieved results were evaluated by performing benchmarks on an industry-related PLC project. The PLC project of the so called "intelligent warehouse" (iStorage), which is shown in Fig. 6, fulfils this criterion.

The iStorage is a part of a flexible production plant that is currently located in the ARENA2036, a research facility at the University of Stuttgart for the next generation of mobility and smart production systems. The system in which the iStorage is integrated produces small model cars out of four different components in total. The intelligent warehouse is responsible for storing those parts and their carriers. Therefore, the system consists of two layers, each including four rows, one for each component type. The lower layer dispenses the different components while the upper layer is responsible for taking back empty carriers. Each row has a capacity for four component carriers. Via light barriers (one for each slot), the system can determine if a carrier is present or not. To prevent the carriers from colliding with each other, three electromagnets are used per row, one between each slot. A magnet is actuated if either its surrounding slots are both occupied or if the second slot in front of it is vacant. A signal light indicates whether actions by a human are necessary, e.g. if each slot of one of the upper rows is occupied or if one component type has run out.

The behavior of the signal light and the rows is controlled by a PLC. In order to fulfill the requirements listed above, the variable list of the PLC consists of one input signal per light barrier and output signals for each magnet and the signal light. This leads to four inputs and three outputs for the row component, whereas the signal light component needs eight inputs and one output. As part of the benchmarks, both components should be synthesized by the demo application, measuring the time needed for the code generation process. Since the inductive code synthesis performed by the tools described in section II.B is non-deterministic, the experiment was repeated ten times in order to calculate the average time and its variance, the latter given by the formula

$$\sigma = \sqrt{\frac{1}{N-1}\sum_{i=1}^{N}(x_i - \bar{x})^2}, \qquad (1)$$

where N stands for the number of repetitions, $x_i$ is the value for the i-th repetition and $\bar{x}$ is the sample average.

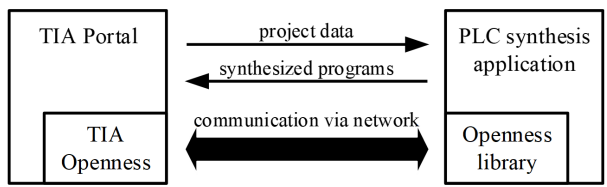

Fig. 4. Communication flows between the TIA portal and synthesis demo application

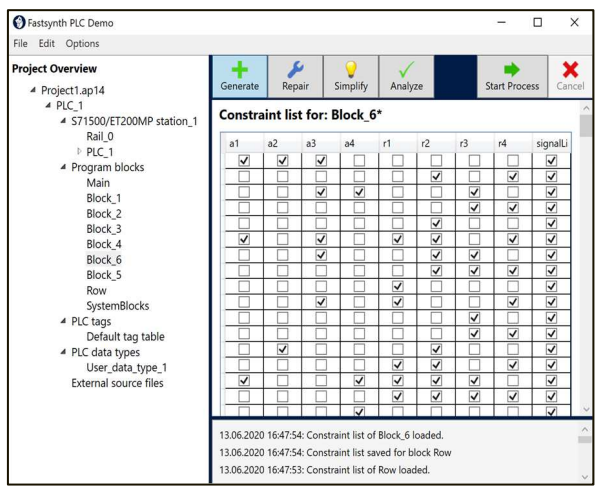

Fig. 5. Main window of the demo application

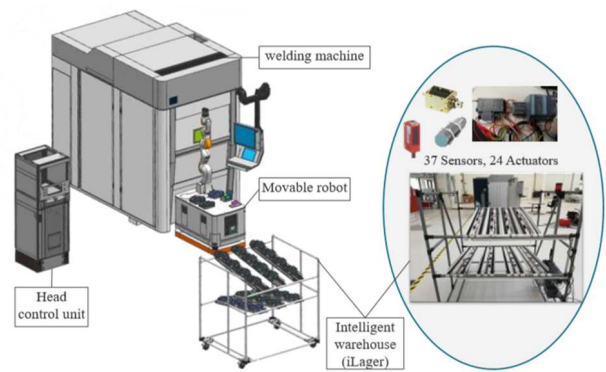

Fig. 6. Flexible production plant with the intelligent warehouse [28]

All benchmark experiments were performed on a personal computer featuring an Intel Core i7-4790K CPU, 16 GB of DDR3 RAM and a Samsung SSD 850 EVO 500GB running Windows 10 Pro (1903).

TABLE I lists the results of the performed benchmark experiments for a single electromagnet of the Row component, all magnets together and the SignalLight component. Overall, it was possible to synthesize both simple and more complex components of the intelligent storage in a time span acceptable to the majority of participants. It is shown that particularly short control commands, as in the case of a single solenoid, can be generated with a delay that is virtually imperceptible to the user. The same applies to the Row module itself, as it simply consists of a series of several solenoid commands. When the synthesis for the block is started, Fastsynth is called once for each output variable, whereby only one solenoid command must be generated for each call. The synthesis time for the Row block therefore increases linearly with the number of solenoids. With the SignalLight module, however, Fastsynth is called only once. The component has more input variables which must be mapped to only one output variable. The result is an exponentially greater solution space and a significantly higher, although still tolerable, synthesis time compared to the other components.

TABLE I. AVERAGE SYNTHESIS TIMES FOR DIFFERENT COMPONENTS OF THE iSTORAGE. 10 TESTS WERE CARRIED OUT PER COMPONENT.

| Component | Average $\bar{x}$ | Std. Deviation $\sigma$ |
|---|---|---|
| Magnet actuation | 128,16 ms | 3,9620 ms |
| Row control | 379,06 ms | 7,9952 ms |
| SignalLight control | 73,125 s | 490,67 ms |

## VI. CONCLUSION

In this paper, the potential of formal methods for code generation and verification in the PLC domain are examined. An engineering-oriented description based upon self-configuring blocks is used to increase accessibility of the underlying concepts in order to facilitate an expert survey. The results of the survey are utilized in a demo application that is integrated into the usual PLC programming workflow using the Siemens TIA Portal. An evaluation is carried out on a discrete manufacturing scenario.

Our main findings are:

- The experts interviewed for this study see great potential for the use of formal methods in the PLC domain. The perceived advantages cover a broad range of use cases and are dependent on the project size and the respective expert's background.
- The integration of formal methods into the PLC programming workflow is possible by extending the TIA Portal through a separate application.
- Code synthesis of low to medium complexity PLC program components can be carried out fast using the proposed approach. The waiting time encountered is considered acceptable by the interviewed experts.

Future work should focus on the possibility of a full integration into TIA as opposed to relying on a separate application. This would improve the accessibility of the proposed method and, thereby, allow for a more widespread utilization.

ACKNOWLEDGMENT

This work was supported by Daniel Kroening of Diffblue Ltd. who took part in the development of CMBC and Fastsynth.

REFERENCES

[1] W. Bolton, *Programmable logic controllers*. Kidlington, Oxford, UK: Newnes, 2015.
[2] S. Biallas, *Verification of Programmable Logic Controller Code using Model Checking and Static Analysis,* 1st ed. Aachen: Shaker, 2016.
[3] *Programmable controllers - Part 3: Programming languages*, IEC 61131-3, 2013.
[4] B. Vogel-Heuser, A. Fay, I. Schaefer, and M. Tichy, "Evolution of software in automated production systems: Challenges and research directions," *Journal of Systems and Software*, vol. 110, pp. 54–84, 2015, DOI: 10.1016/j.jss.2015.08.026.
[5] K. Schneider, *Verification of reactive systems: Formal methods and algorithms*. Berlin, Heidelberg: Springer, 2004.
[6] D. Basile, M. H. ter Beek, A. Fantechi, S. Gnesi, F. Mazzanti, A. Piattino, D. Trentini, and A. Ferrari, "On the Industrial Uptake of Formal Methods in the Railway Domain," in *Lecture Notes in Computer Science, INTEGRATED FORMAL METHODS: 14th international conference, ifm 2018, maynooth*, C. A. Furia and K. Winter, Eds., [Place of publication not identified]: SPRINGER INTERNATIONAL PU, 2018, pp. 20–29, DOI: 10.1007/978-3-319-98938-9_2.
[7] V. Wiels, R. Delmas, D. Doose, P. L. Garoche, J. Cazin, and G. Durrieu, "Formal Verification of Critical Aerospace Software," *Journal AerospaceLab*, vol. 10, no. 4, pp. 1–8, 2012.
[8] O. Niggemann, A. Maier, and Jasperneite Juergen, "Model-based Development of Automation Systems," in *Tagungsband Modellbasierte Entwicklung eingebetteter Systeme*.
[9] A. Fay, B. Vogel-Heuser, T. Frank, K. Eckert, T. Hadlich, and C. Diedrich, "Enhancing a model-based engineering approach for distributed manufacturing automation systems with characteristics and design patterns," *Journal of Systems and Software*, vol. 101, pp. 221–235, 2015, DOI: 10.1016/j.jss.2014.12.028.
[10] F. Rademacher, S. Sachweh, and A. Zundorf, "Differences between Model-Driven Development of Service-Oriented and Microservice Architecture," in *2017 IEEE International Conference on Software Architecture Workshops (ICSAW)*, Gothenburg, Sweden, Apr. 2017 - Apr. 2017, pp. 38–45, DOI: 10.1109/ICSAW.2017.32.
[11] M. F. Zaeh and C. Poernbacher, "Model-driven development of PLC software for machine tools," *Prod. Eng. Res. Devel.*, vol. 2, no. 1, pp. 39–46, 2008, DOI: 10.1007/s11740-008-0083-7.
[12] M. E. Witte, C. Diedrich, and H. Figalist, "Model-based development in automation," *at - Automatisierungstechnik*, vol. 66, no. 5, pp. 360–371, 2018, DOI: 10.1515/auto-2017-0125.
[13] J. Woodcock, P. G. Larsen, J. Bicarregui, and J. Fitzgerald, "Formal methods," *ACM Comput. Surv.*, vol. 41, no. 4, pp. 1–36, 2009, DOI: 10.1145/1592434.1592436.
[14] J. R. Kiniry and F. Fairmichael, "Ensuring Consistency between Designs, Documentation, Formal Specifications, and

Implementations," in *Lecture Notes in Computer Science, Component-Based Software Engineering*, G. A. Lewis, I. Poernomo, and C. Hofmeister, Eds., Berlin, Heidelberg: Springer Berlin Heidelberg, 2009, pp. 242–261, DOI: 10.1007/978-3-642-02414-6_15.

[15] G. Bahig and A. El-Kadi, "Formal Verification of Automotive Design in Compliance With ISO 26262 Design Verification Guidelines," *IEEE Access*, vol. 5, pp. 4505–4516, 2017, DOI: 10.1109/ACCESS.2017.2683508.

[16] *Road vehicles – Functional safety*, ISO 26262, 2018.

[17] D. Darvas, I. Majzik, and E. Blanco Viñuela, "Formal Verification of Safety PLC Based Control Software," in *Lecture Notes in Computer Science, Integrated Formal Methods*, E. Ábrahám and M. Huisman, Eds., Cham: Springer International Publishing, 2016, pp. 508–522, DOI: 10.1007/978-3-319-33693-0_32.

[18] G. Frey and L. Litz, "Formal methods in PLC programming," in *SMC 2000 Conference Proceedings. 2000 IEEE International Conference on Systems, Man and Cybernetics. 'Cybernetics Evolving to Systems, Humans, Organizations, and their Complex Interactions' (Cat. No.00CH37166)*, Nashville, TN, USA, 2000, pp. 2431–2436, DOI: 10.1109/ICSMC.2000.884356.

[19] O. Ljungkrantz, K. Akesson, M. Fabian, and C. Yuan, "A formal specification language for PLC-based control logic," in *2010 8th IEEE International Conference on Industrial Informatics*, Osaka, Japan, 072010, pp. 1067–1072, DOI: 10.1109/INDIN.2010.5549591.

[20] D. Darvas, E. B. Vinuela, and I. Majzik, "PLC code generation based on a formal specification language," in *2016 IEEE 14th International Conference on Industrial Informatics (INDIN)*, Poitiers, France, 72016, pp. 389–396, DOI: 10.1109/INDIN.2016.7819191.

[21] H. Barbosa and D. Déharbe, "Formal Verification of PLC Programs Using the B Method," in *Lecture Notes in Computer Science, Abstract State Machines, Alloy, B, VDM, and Z*, D. Hutchison et al., Eds., Berlin, Heidelberg: Springer Berlin Heidelberg, 2012, pp. 353–356, DOI: 10.1007/978-3-642-30885-7_30.

[22] A. Souri and M. Norouzi, "A State-of-the-Art Survey on Formal Verification of the Internet of Things Applications," *J Serv Sci Res*, vol. 11, no. 1, pp. 47–67, 2019, DOI: 10.1007/s12927-019-0003-8.

[23] Diffblue Ltd., *cbmc*. [Online] Available: https://github.com/diffblue/cbmc/tree/5d6d1aead28a1c8b2bbb54ddd51e30a51770ea9c. Accessed on: Nov. 30 2019.

[24] E. Clarke, D. Kroening, and F. Lerda, "A Tool for Checking ANSI-C Programs," in *Tools and Algorithms for the Construction and Analysis of Systems (TACAS 2004)*, 2004, pp. 168–176.

[25] Siemens, *Programmierleitfaden für S7-1200/S7-1500*. [Online] Available: https://www.automation.siemens.com/sce-static/learning-training-documents/tia-portal/hw-config-s7-1200/programming-guideline-de.pdf. Accessed on: Nov. 27 2019.

[26] A. Abate, C. David, P. Kesseli, D. Kroening, and E. Polgreen, "Counterexample Guided Inductive Synthesis Modulo Theories," in *Computer Aided Verification (CAV)*, 2018, pp. 270–288.

[27] Diffblue Ltd., *fastsynth*. [Online] Available: https://github.com/kroening/fastsynth/tree/91cc3c01ef983a60f2a7c8cc312c8e3525597778. Accessed on: Dec. 01 2019.

[28] Behrang Ashtari Talkhestani, D. Braun, W. Schloegl, and M. Weyrich, "Qualitative and quantitative evaluation of reconfiguring an automation system using Digital Twin," 2020, DOI: 10.1016/j.procir.2020.03.014.